\documentclass[12pt,twoside]{article}
\usepackage{amsmath,amssymb}

\let\lsim\lesssim
\let\gsim\gtrsim
\let\overdot\dot

\parskip=\smallskipamount
\begin{document}

\title{Four Big Questions with Pretty Good Answers}
\author{Frank Wilczek\\
\small\it Center for Theoretical Physics\\ 
\small\it Massachusetts Institute of Technology\\ 
\small\it Cambridge, MA 02139-4307}
\date{\small MIT-CTP \# 3236\\[1ex]
Talk given at a
Symposium in Honor of Heisenberg's 100th birthday, December 6, 2001,  Munich. 
To be published in the Festschrift.}

\maketitle 

\markboth{Frank Wilczek}{Heisenberg -- Four Big Questions}

\begin{abstract}\noindent I discuss four big questions that can be importantly addressed
using concepts from modern QCD.   They concern the origin of mass, the feebleness of
gravity, the uniqueness of physical laws, and the fate of matter when it is squeezed
very hard.  
\end{abstract}

\thispagestyle{empty}
\pagebreak

Heisenberg's motivation for studying physics was not only to solve particular
problems, but also to illuminate the discussion of broad philosophical questions. 
Following his epochal contribution, at a very young age, to the foundation of
quantum physics,  most of Heisenberg's scientific life was devoted to searching for
the fundamental laws underlying nuclear physics.    In celebrating the one
hundredth anniversary of his birth,  I think it is appropriate to consider how our
decisive progress in uncovering those laws has advanced the discussion of some
quite basic -- you might call them either ``big'' or ``naive'' -- questions about
Nature.    These are insights I'd like to share with Heisenberg if he could be present
today.  I think he'd enjoy them.

\section{What is the Origin of Mass?}

\subsection{Framing the Question}

That a question makes grammatical sense does not guarantee that it is answerable,
or even coherent.  Indeed, this observation is a central theme of Heisenberg's early,
classic exposition of quantum theory \cite{heisenberg}.  In that spirit, let us begin with a critical 
examination of the question posed in this Section: What is the origin of mass? \cite{section1}  

In classical mechanics mass appears as a
primary concept.   It was a very great step for the founders of classical mechanics to
isolate the scientific concept of mass.   In Newton's laws of motion, mass appears as an
irreducible, intrinsic property of matter, which relates its manifest response
(acceleration) to an abstract cause (force).    An object without mass would not know
how to move.  It would not know, from one moment to the next, where in space it was
supposed to be.  It would be, in a very strong sense, unphysical.   Also, 
in Newton's law of gravity, the mass of an object
governs the strength of the force it exerts.  One cannot
build up an object that gravitates,  out of material that does not.  Thus it is difficult to
imagine, in the Newtonian framework, what could possibly constitute an ``origin of
mass''.    In that framework, mass just is what it is.  

Later developments in physics make the concept of mass seem less
irreducible.   The undermining process started in earnest with the theories of
relativity.    The famous equation $E = mc^2$ of special relativity theory, written
that way, betrays the prejudice that we should express energy in terms of mass.   But
we can also read it as $m= E/c^2$, which  suggests the possibility of explaining mass
in terms of energy.   In general relativity the response of matter to gravity is
independent of mass (equivalence principle), while space-time curvature is
generated directly by energy-momentum, according to 
$R_{\mu\nu} - \frac12 g_{\mu\nu} R = \kappa T_{\mu\nu}$, with
$\kappa \equiv 8\pi G_N /c^2$.   Mass appears as a contributing factor to
energy-momentum, but it has no uniquely privileged status.   

At an abstract level, mass appears as a label for irreducible representations of the
Poincare group.  Since representations with $m\neq 0$ appear in tensor products of
$m=0$ representations it is possible, at least kinematically, to build massive particles
as composites of massless particles, or massless particles and fields.   

\subsubsection{Lorentz's Dream}

At a much more concrete level, the question of the origin of mass virtually forced
itself upon physicists' attention in connection with the development of electron
theory.   Electrons generate electromagnetic fields; these fields have energy and
therefore inertia.  Indeed,  a classical point electron is surrounded by an electric
field varying as
${e}/{r^2}$.   The energy in this field is infinite, due to a divergent contribution
around $r
\rightarrow 0$.   It was a dream of Lorentz (pursued in evolved forms by many
others including Poincare, Dirac, Wheeler, and Feynman), to account for the
electron's mass entirely in terms of its electromagnetic fields, by using a more
refined picture of electrons.  Lorentz hoped that in a correct model of electrons they
would emerge as extended objects,  and that the energy in the Coulomb field would
come out finite, and  account for all (or most) of the inertia of electrons.  

Later progress in the quantum theory of electrons rendered this program moot by
showing that the charge of an electron,  and therefore of course its associated
electric field,  is intrinsically smeared out by quantum fluctuations in its position.  
Indeed, due to the uncertainty principle the picture of electrons as ideal point
particles certainly breaks down for distances $r \lsim
\hbar/mc$, the Compton radius.  At momenta $p \gsim \hbar/r$, the velocity
$p/m$ formally becomes of order
$c$, and one cannot regard the electron as a static point source.   If we cut off the
simple electrostatic calculation at the Compton radius, we find an electromagnetic
contribution to the electron mass of order
$\delta m \sim \alpha m$, where
$\alpha = e^2/4\pi\hbar c
\approx 1/137$ is the fine structure constant.   In this sense the easily identifiable
and intuitive electromagnetic contribution to the mass, which Lorentz hoped to
build upon, is small.  To go further, we cannot avoid considering relativity and
quantum mechanics together.    That means quantum field theory.  

\subsubsection{Its Debacle}

In quantum electrodynamics itself, the whole issue of the electromagnetic
contribution to the electron mass becomes quite dodgy, due to renormalization.  
Quantum electrodynamics does not exist nonperturbatively.  One can regulate and
renormalize order-by-order in perturbation theory, but there are strong arguments
that the series does not converge, or even represent the asymptotic expansion of a
satisfactory theory.  In a renormalization group analysis, this is because the effective
coupling blows up logarithmically at short distances, and one cannot remove the
cutoff.  In a lattice regularization, one could not achieve a Lorentz-invariant
limit.\footnote{Actually this blow-up, the famous  Landau pole, arises from
extrapolating the perturbative result beyond its range of validity.   What one can
deduce simply and rigorously is that the effective coupling does not become small at
short distances: QED is not asymptotically free.    If there is a fixed point at finite
coupling, it may be possible to construct a relativistically invariant limiting theory. 
But even if such a theory were to exist, its physical relevance would be quite
dubious, since we know that there's much more to physics than electrodynamics at
distances so short that the logarithms matter.}   So one cannot strictly separate the
issue of electromagnetic mass from the unknown physics that ultimately regularizes
the short-distance singularities of QED.   Perturbatively, the mass is multiplicatively
renormalized, by a factor that diverges as the regulator is removed.    Since results
for different values of the charge are incommensurate, one does not obtain a
well-defined,  finite answer for the electromagnetic contribution to the mass.  If we
regard QED as an effective theory, essentially by leaving in an energy cutoff
$\Lambda$, corresponding to a distance cutoff $\hbar c/\Lambda$, we get a
contribution to the mass at short distances going as
$\delta m \propto \alpha m\log (\Lambda/m)$.  Quantum mechanics has changed
the power-law divergence into a logarithm.  As a result, $\delta m$ is a fractionally
small contribution to the total mass, at least for sub-Planckian 
$\Lambda$ (i.e., $\Lambda
\lsim 10^{19}$ GeV).  We know that QED ceases to be a complete description of
physics, or even a well-isolated theory, far below such energies.   In any case, since
the mass renormalization is multiplicative, a zero-mass electron stays at zero mass.  
Indeed, the massless version of QED has enhanced symmetry -- chiral symmetry --  
which is not destroyed by (perturbative) renormalization.    

In short, very little seems to survive from Lorentz's original dream. I've described
this fiasco in some detail, since it provides an instructive background to contrast
with our upcoming considerations.  

\subsubsection{Upping the Ante}

Quantum field theory completely changes how we view the
question of the origin of mass.  As we have seen, it quashes hope for a simple,
mechanistic classical explanation.   At a more profound level  it makes the
question seem much more central.  Indeed, renormalizable quantum field theories
are extremely restricted.  They tend to contain few parameters (unless of course they
contain many fields and few symmetries), among which masses feature
prominently.   Moreover, they feature  enhanced symmetry when mass
parameters vanish.   So massless theories are significantly more constrained, and
there is a definite sense in which they are prettier.     These motivations  survive,
even if one is not committed to renormalizability.

\subsection{Most of the Answer: QCD Lite}

Enough of generalities!  I want now to describe some very beautiful and specific
insights into the origin of mass in the real world.   We will construct -- following
Nature -- mass without mass, using only $c$ and $\hbar$.  

\subsubsection{Introducing QCD Lite}

My central points are most easily made with reference to a slight idealization of QCD
which I call, for reasons that will be obvious, QCD Lite.  It is a  nonabelian gauge
theory based on the gauge group $SU(3)$ coupled to two triplets and two
anti-triplets of left-handed fermions, all with zero mass.   Of course I have in mind
that the gauge group represents color, and that one set of triplet and antitriplet will
be identified with the quark fields $u_L, u_R$ and the other with
$d_L, d_R$.  

Upon demanding renormalizability,\footnote{Or, in physical terms, the existence of
a relativistically invariant limiting theory.  Or alternatively, radical decoupling from
an unspecified, gratuitous high-energy cutoff.} this theory appears to contain
precisely one parameter, the coupling $g$.   It is, in units with $\hbar = c=1$, a pure
number.    I'm ignoring the $\theta$ parameter, which has no physical content here,
since it can be absorbed into the definition of the quark fields.   Mass terms for the
gluons are forbidden by gauge invariance.  Mass terms for the quarks are forbidden
by chiral $SU(2)_L \times SU(2)_R$ flavor symmetry.  

\subsubsection{Running Coupling; Dimensional Transmutation}

The coupling constant $g$ that appears in the Lagrangian of QCD Lite, like the
corresponding constant $e$ in QED,  is a dimensionless number (in units with
$\hbar = c = 1$).   Likewise for the fine-structure constant $\alpha_s \equiv
g^2/4\pi$.   But the real situation, when we take into account the effect of quantum
mechanics, is quite different.   Empty space is a medium, full of virtual particles, and
responds dynamically to charges placed within it.  It can be polarized, and the
polarization clouds surrounding test charges can shield (screen) or enhance
(antiscreen) their strength.  In other words,  quantum-mechanically the measured
strength of the coupling depends on the distance scale, or equivalently the (inverse)
energy, scale at which it is measured: $\alpha_s \rightarrow \alpha_s(Q)$.   This
is a central feature of QCD Lite, and of course of QCD itself.    These
theories predict that the effective coupling gets small at large $Q$, or equivalently at
short distance.   

This behavior displays itself in a remarkably direct and tangible form in the final
states of electron-positron annihilation.     Hadrons emerging from high-energy
electron-positron annihilation organize themselves into collimated jets.  Usually
there are two jets, but occasionally three.   The theoretical interpretation is profound
but, given asymptotic freedom, straightforward.  The primary products emerging
from the annihilation are a quark and an antiquark.  They emit soft -- that is, low
energy-momentum -- radiation copiously, but only rarely hard radiation.  That's a
restatement, in momentum space, of asymptotic freedom.  The soft radiation
materializes as many particles, but these particles inherit their direction of flow from
the quark or antiquark, and therefore constitute a jet.   In the relatively rare case
that there is hard radiation, that is to say emission of an energetic gluon, the gluon
induces its own independent jet.    All this can be made completely quantitative.  
There are precise predictions for the ratio of three- to two-jet events, the rare
occurrence of four- or more jet events, how these ratios change with energy, angular
dependence, and so forth.  The observations agree with these predictions.   Thus
they provide overwhelming, direct evidence for the most basic elements of the
theory, that is the quark-gluon and gluon-gluon couplings.

Because the coupling runs we can, within any given version of QCD Lite, measure
any given {\it numerical\/} value 
$a=\alpha_s(Q)$, simply by choosing an appropriate
$Q$.  It appeared, classically, that we had an infinite number of different versions of
QCD Lite, with different values of the coupling parameter.  In reality the only
difference among all these theories, after they are quantized, is the unit they choose
for measuring mass.  All dimensionless physical parameters, and in particular all
mass ratios, are uniquely determined.   We can trade the dimensionless parameter
$g$ for the unit of mass.   This is the phenomenon of dimensional transmutation.

Of course the value of the overall energy scale makes a big difference when we come
to couple QCD  Lite, or of course QCD, to the rest of physics.  Gravity, for example,
cares very much about the absolute value of masses.   But within QCD Lite itself, if 
we compute any dimensionless quantity whatsoever, we will obtain a unique answer,
independent of any choice of coupling parameter.     Thus, properly understood, the
value of the QCD coupling constant does not so much govern QCD itself -- within its
own domain, QCD is essentially unique --  but  rather how QCD fits in with the rest of
physics.

\subsubsection{Physical Mass Spectrum -- QCD Lite and Reality}

Now let us consider more concretely how these dynamical phenomena lead us to a
non-trivial hadron spectrum. Looking at the classical equations of  QCD,  one would
expect an attractive force between quarks that varies with the distance as
$g^2/4\pi r^2$,  where $g$ is the coupling constant.  This result is modified, however, by
the effects of quantum fluctuations.   As we have just discussed, the omnipresent evanescence of virtual
particles renders empty space into a dynamical medium,  whose response alters the
force law.   The antiscreening effect of virtual color gluons (asymptotic
freedom), enhances the strength of the attraction, by a factor which grows with the
distance.     This effect can be captured by defining an effective coupling, $g(r)$,
that grows with distance.  

The attractive interaction among quarks wants to bind
them together;  but the potential energy to be gained by bringing quarks together
must be weighed against its cost in kinetic energy.    In a more familiar application,
just this sort of competition between Coulomb attraction and localization energy is
responsible for the stability and finite size of atoms.   Here, quantum-mechanical
uncertainty implies that quark wave-functions localized in space must contain a
substantial admixture of high momentum.  For a relativistic particle, this translates
directly into energy.      If the attraction followed Coulomb's law, with a small
coupling, the energetic price for staying localized would always outweigh the profit
from attraction, and the quarks would not form a bound state.  
Indeed, the kinetic energy $\hbar c/r$ beats the potential energy
$g^2/4\pi r$.   But the running
coupling of QCD grows with distance, and that tips the balance.   The quarks finally
get reined in, at distances where $\alpha_s(r)$ becomes large.  

We need not rely on heuristic pictures, or wishful thinking, to speculate about the
mass spectrum of QCD Lite.  It has been calculated by direct numerical integration of
the fundamental equations, using the techniques of lattice gauge
theory\footnote{There are significant technical issues around realizing chiral
symmetry in numerical work involving discretization on a  lattice.  Recent
theoretical work appears to have resolved the conceptual issues, but the numerical
work does not yet fully reflect this progress.   To avoid a fussy presentation I've
oversimplified by passing over these issues, which do not affect my main point.}. 
The results bear a remarkable qualitative and semi-quantitative resemblance to the
observed spectrum of non-strange hadrons, generally at the 10\% level,
comfortably within known sources of error due to finite size, statistics, etc. -- and
(anticipating) small quark masses.  Of course, in line with our preceding discussion,
the overall {\it scale\/} of hadron masses is not determined by the theory.   But all
mass ratios are predicted, with no free parameters, as of course are the resonance
quantum numbers.   

QCD Lite is not the real world, of course.   So although in QCD Lite we get mass
without mass in the strict sense, to assess how much real-world mass arises this way,
we need to assess how good an approximation QCD Lite is to reality, quantitatively. 
We can do this by adjusting the non-zero values of $m_u$ and
$m_d$ to make the spectrum best fit reality, and then seeing how much they
contributed to the fit.\footnote{Again, there are significant technical issues here,
especially regarding the role of the strange quark.  Fortunately, the uncertainties are
numerically small.}   Unlike charges in QED, masses in QCD are soft perturbations, and we
can calculate a meaningful finite difference between the spectra of these two
theories.  There is also a well-developed alternative approach to estimating the
contribution of quark masses, by exploiting the phenomenology of chiral symmetry
breaking.   Either way, one finds that the quark masses contribute at most a few per
cent to the masses of protons and neutrons.  

Protons and neutrons, in turn, contribute more than 99\% of the mass of ordinary
matter.  So QCD Lite provides, for our purpose, an excellent description of reality.   
The origin of the bulk of the mass of ordinary matter is well accounted for, in a
theory based on pure concepts and using no mass parameters -- indeed, no mass {\it
unit\/} --  at all!

\subsubsection{Comparing With the Old Dream}

While our final result realizes something very close to Lorentz's dream, the details
and the mechanism are quite different.   

Obviously we are speaking of hadrons, not electrons, and of QCD, not classical
electrodynamics.  The deepest difference, however, concerns the source and location
of the energy whereby
$m=E/c^2$ is realized.   In Lorentz's dream, the energy was self-energy, close to the
location of the point particle.  In QCD Lite the self-mass vanishes.   Paradoxically,
there is a sense in which the self-energy of a quark\footnote{Infinite self-energy
does {\it not\/} conflict with zero mass. 
$E=mc^2$ describes the energy of a particle of mass $m$ when it is at rest; but of
course, as we know from photons, there can also be energy in massless particles,
which cannot be brought to rest.} is infinite (confinement), but this is due to the
spatial extent of its color field, which has a tail extending to infinity, not to any
short-distance singularity.  To make physical hadrons, quarks and gluons must be
brought together, in such a way that the total color vanishes.  Then there is no
infinite tail of color flux;  the different tails have cancelled. But at finite distances the
cancellation is incomplete, because Heisenberg's uncertainty principle imposes an
energetic cost for keeping color charges precisely localized together.   The bulk of
the mass of the hadrons comes from the residues of these long tails, not from
singularities near point-like color charges.   

\subsection{(Many) Remaining Issues}

While the dynamical energy of massless QCD accounts for the bulk of mass, for
ordinary matter, it is far from being the only source of mass in Nature.   

Mass terms for quarks and charged leptons appear to violate the electroweak gauge
symmetry
$SU(2)\times U(1)$.   But gauge symmetry cannot be violated in the fundamental
equations -- that would lead to ghosts and/or non-unitarity, and prevent
construction of a sensible quantum theory.  So these masses must, in a sense,  have
their ``origin'' in spontaneous symmetry breaking.  That is accomplished, in the
Standard Model, by having a non-singlet Higgs field acquire a vacuum expectation
value.    Why this value is so small, compared to the Planck scale, is one aspect of
what is usually called the hierarchy problem.   Why the couplings of this field are so
disparate -- especially, what is particularly crucial to the structure of physical
reality,  why its dimensionless couplings to $e, u, d$ are so tiny (in the range
~$10^{-5}-10^{-6}$) -- is an aspect of what is usually called the flavor problem.  

Then there are separate problems for generating masses of supersymmetric particles
(soft breaking parameters, $\mu$ term), for generating the mass of cosmological
`dark matter' (this might be included in the previous item!), for generating neutrino
masses, and apparently for generating the mass density of empty space
(cosmological term).   

Obviously, many big questions about the origin of mass remain.    But I think we've
answered a major one beautifully and convincingly.

\section{Why is Gravity Feeble?}

Gravity dominates the large-scale structure of the Universe, but only so to speak by
default \cite{section2}.   Matter arranges itself to cancel electromagnetism, and the strong and weak
forces are intrinsically short-ranged.    At a more fundamental level, gravity is
extravagantly feeble.    Acting between protons, gravitational attraction is about
$10^{-36}$ times weaker than  electrical repulsion.    Where does this outlandish
disparity from?  What does it mean?  

Feynman wrote
\begin{quote} There's a certain irrationality to any work on [quantum] gravitation,
so it's hard to explain why you do any of it ...  It is therefore clear that the problem
we working on is not the correct problem;  the correct problem is  What determines
the size of gravitation?
\end{quote}

I want to argue that today it is natural to see the problem of why gravity is
extravagantly feeble  in a new way -- upside-down and through a distorting lens
compared to its superficial appearance.   When viewed this way, it comes to seem
much less enigmatic.  

First let me quantify the problem.   The mass of ordinary matter is dominated by
protons (and neutrons), and the force of gravity is proportional to (mass)$^2$.   
From Newton's constant, the proton mass, and fundamental constants we can form
the pure dimensionless number
$$ X~=~  G_N m_p^2 /\hbar c~,
$$ where $G_N$ is Newton's constant, $m_p$ is the proton mass, $\hbar$ is
Planck's constant, and $c$ is the speed of light.    Substituting the measured values,
we obtain
$$ X~\approx~ 6\times 10^{-39} ~.
$$ This is what we mean, quantitatively, when we say that gravity is extravagantly
feeble.   

We can interpret $X$ directly in physical terms, too.    Since the proton's
geometrical size $R$ is roughly the same as its Compton radius $\hbar/m_pc$,  the
gravitational binding energy of a proton is roughly $G_N m_p^2/ R \approx  X m_p
c^2$.   So $X$ is the fractional contribution of gravitational binding energy to the
proton's rest mass!

\subsubsection{Planck's Astonishing Hypothesis}

An ultimate goal of physical theory is to explain the world purely conceptually, with
no parameters at all.   Superficially, this idea seems to run afoul of dimensional
analysis -- concepts don't have units, but physical quantities do!   

There is a sensible
version of this goal, however, that is rapidly becoming conventional wisdom, despite
falling well short of scientific knowledge.  Soon after he introduced his constant
$\hbar$, in the course of a phenomenological fit to the black-body radiation
spectrum,  Planck pointed out the possibility of building a system of units based on
the three fundamental constants $\hbar, c, G_N$.    Indeed, from these three we can
construct a unit of mass $(\hbar c /G_N)^{1/2}$, a unit of of length
$(\hbar G_N/c^3)^{1/2}$, and a unit of time $(\hbar G_N/c^5)^{1/2}$ --
what we now call the Planck mass, length, and time respectively.    

Planck's proposal for a system of units based on fundamental physical constants was,
when it was made, rather thinly rooted in physics.    But by now there are profound
reasons to regard $c$,
$\hbar$ and $G$ as  {\it conversion factors\/} rather than  numerical parameters.
In the special theory of relativity,  there are symmetries relating space and time --
and $c$ serves as a conversion factor between the units in which space-intervals and
time-intervals are measured.   In quantum theory,  the energy of a state is
proportional to the frequency of its oscillations -- and
$\hbar$  is the conversion factor.    Thus $c$ and $\hbar$ appear directly as
measures in the basic laws of these two great theories.    Finally, in general relativity
theory, space-time curvature is proportional to the density of energy -- and $G_N$
(actually $G_N/ c^4$) is the conversion factor.     

If we want to adopt Planck's astonishing hypothesis, that we must build up physics
solely from these three conversion factors, then the enigma of $X$'s smallness looks
quite different.  We see that the question it poses is not``Why is gravity so feeble?''  
but rather ``Why is the proton's mass so small?".   For according to Planck's
hypothesis, in natural (Planck) units the strength of gravity simply is what it is, a
primary quantity.  So it can only be the proton's mass which provides the tiny
number
$\sqrt X$.  

\subsubsection{Running in Place} 

That's a provocative and fruitful way to invert the question, because we now have
a quite  deep understanding of the origin of the proton's mass,  as I've just
reviewed.  

The proton mass is determined, according to the dynamics I've described,  by the
distance at which the running QCD coupling becomes strong.  Let's call this the
QCD-distance.    Our question, ``Why is the proton mass  so small?'' has been
transformed into the question, ``Why is the QCD-distance is much larger than the
Planck length?''   To close our circle of ideas we need to explain,   if only the Planck
length is truly fundamental,  how it is that such a vastly different length arises
naturally.   

This last elucidation, profound and beautiful, is worthy of the problem.   It has to do
with how the coupling runs, in detail.   When the QCD coupling is weak, `running' is
actually a bit of a misnomer.   Rather, the coupling creeps along like a wounded
snail.  We can in fact calculate the behavior precisely, following the rules of
quantum field theory, and even test it experimentally, as I mentioned before.  The
inverse coupling varies logarithmically with distance.   Therefore, if we want to
evolve an even moderately small coupling into a coupling of order unity, we must
let it between length-scales whose ratio is exponentially large.    So if the QCD
coupling is even moderately small at the Planck length, assumed fundamental,  it
will only reach unity at a much larger distance.   

Numerically, what we predict is that $\alpha_s(l_{\rm Pl.})$ at the Planck length is
roughly a third to a fourth of  what it is observed to be at
$10^{-15}$~cm; that is, $\alpha_s(l_{\rm Pl.}) \approx 1/30$.   We cannot measure
$\alpha_s(l_{\rm Pl.})$ directly, of course, but there are good independent reasons, 
having to do with the unification of couplings, to believe that this value holds in
reality.  It is amusing to note that in terms of the coupling itself, what we require is  
$g_s(l_{\rm Pl.}) \approx 1/2$!  From this modest and seemingly innocuous
hypothesis, which involves neither really big numbers nor speculative dynamics
going beyond what is supported by hard experimental evidence, we have produced a
logical explanation of the tiny value of $X$.

\section{Are the Laws of Physics Unique?}

This will be by far the most speculative portion of my talk.  I think the interest, and
the difficulty, of the question justifies a liberal standard.  {\it Caveat emptor}.

\subsection{Interpreting the Question}

Prior to the twentieth century, in classical physics, there was a clear separation
between dynamical equations and initial conditions.  The dynamical equations could
predict, given the state of matter at one time, its behavior in the future.   But these
equations did not say much about the specific forms in which matter actually exists. 
In particular, there was no explanation of why there should be such a subject as
chemistry, let alone its content, nor of the origin of astronomical objects.   One
could, and as far as I know essentially everyone did, assume that the laws are
universally valid without feeling  burdened to explain every specific feature of the
actual Universe.  

Over the past century the situation changed qualitatively.   Developments in
quantum theory, starting with the Bohr atom and culminating in the Standard
Model, give us a remarkably complete and accurate description of the basic
interactions of ordinary matter.   I don't think many physicists doubt that this
description is sufficient, {\it in principle}, to describe its specific forms.  (As a
practical matter, of course, our ability to exploit the equations is quite limited.)
Developments in cosmology have revealed an amazing uniformity and simplicity to
the large-scale structure of the Universe, and allowed us to sketch a plausible
account of origins starting with a very limited set of input parameters from the early
moments of the Big Bang.

Having come so far, we can begin to wonder whether, or in what sense, it is possible
to go all the way.  Is it possible to explain (in principle) everything about the
observed Universe from the laws of physics?   As our friends in biology or history
would be quick to remind us, this question is quite pretentious and ridiculous, at a
very basic level.   The laws of physics are never going to allow  you to derive, even in
principle, the mechanism of the cell cycle, the rhetoric of the Gettysburg address, or
indeed the vast bulk of what is of interest in these subjects.     Closer to home, it
would appear that the specific number and placement of planets in the Solar System,
for which Kepler hypothesized a unique mathematical explanation involving regular
solids, is accidental.   Indeed, recently discovered extra-solar planetary systems, not
to mention the system of Jupiter's moons, have quite different sizes and shapes.   

It is conceivable, I suppose, that all these differences could arise from our limited
perspective for viewing the quantum-mechanical wave function of the entire
Universe, which itself is uniquely determined.  In this conception, the accidents of
history would be a matter of which branch of the wave-function we happen to live
on.  They would be functions depending on which particular one among the `many
worlds' contained in the Universal wave-function we happen to inhabit.    The
question whether physics could explain everything would still be pretentious and
ridiculous, but its answer might be weirdly satisfying.  Physics would explain a great
many things, and also explain why it could not explain the other things.   

In any case, it is perfectly clear that there are an enormous number of facts about
the Universe that we will not be able to derive from a cohesive framework of
generally applicable laws of physics.    We comfort ourselves by giving them a name,
contingent facts, with the connotation that they might have been different.    With
this background, it is natural to interpret the question that entitles this Section in a
more precise way, as follows.  Are there contingent regularities of the whole
observable Universe?  If so, there is a definite sense in which the laws of physics are
not unique.  I will now show, in the context of a reasonably orthodox world-model,
how it could be so.

\subsection{A Model of World Non-Uniqueness}

\subsubsection{Relevant Properties of Axions}

I will need to use a few properties of axions, which I should briefly recall \cite{axionreviews}.  

Given its extensive symmetry and the tight structure of relativistic quantum field
theory, the definition of QCD only requires, and only permits, a very restricted set of
parameters.  These consist of the coupling constant and the quark masses, which
we've already discussed, and one more -- the so-called
$\theta$~parameter.  Physical results depend periodically upon $\theta$, so that
effectively it can take values between $\pm \pi$.  We don't know the actual value of
the
$\theta$ parameter, but only a limit, $|\theta | \lsim 10^{-9}$.   Values outside this
small range are excluded by experimental results, principally the tight bound on the
electric dipole moment of the neutron.   The discrete symmetries P and T are
violated by  $\theta$ unless $\theta \equiv 0$ (mod $~\pi$).   Since there are P and
T violating interactions in the world, the $\theta$ parameter cannot be put to zero
by any strict symmetry assumption.   So its smallness is a challenge to understand.  

The effective value of $\theta$ will be affected by dynamics, and in particular by
condensations (spontaneous symmetry breaking).   Peccei and Quinn discovered
that if one imposed a certain asymptotic symmetry, and if that symmetry were
spontaneously broken, then an effective value
$\theta \approx 0$ would be obtained.  Weinberg and I explained that the approach
$\theta
\rightarrow 0$ could be understood as a relaxation process, wherein a very light
collective field, corresponding quite directly to $\theta$, settled down to its
minimum energy state.  This is the axion field, and its quanta are called axions.  

The phenomenology of axions is essentially controlled by one parameter, $F$.  $F$
has dimensions of mass.  It is the scale at which Peccei-Quinn symmetry breaks. 
More specifically, there is some scalar field $\phi$ that carries Peccei-Quinn charge
and acquires a vacuum expectation value of order $F$.  (If there are several
condensates, the one with the largest vacuum expectation value dominates.)  The potential
for $| \phi |$ can be complicated and might involve very high-scale physics, but the
essence of Peccei-Quinn symmetry is to posit that the classical Lagrangian is
independent of the phase of $\phi$, so that the only way in which that phase affects
the theory is to modulate the effective value of the
$\theta$ term, in the form $\theta_{\rm eff.} = \theta_{\rm bare} + ~{\rm
arg}~\phi$.\footnote{I am putting a standard integer-valued parameter, not discussed here, 
$N=1$, and slighting several other inessential
technicalities.}  Then we identify the axion field
$a$ according to
$\langle \phi \rangle \equiv  F e^{ia/F}e^{-i\theta_{\rm bare}}$, so $\theta_{\rm eff.} = 
a/F$.   This insures canonical normalization of the kinetic energy for
$a$.  

In a crude approximation, imagining weak coupling, the potential for $a$ arises
from instanton and anti-instanton contribution, and takes the form
${\frac12}(1-\cos \theta_{\rm eff.})\times\linebreak
e^{{-8\pi^2}/{g^2}}\Lambda_{\rm QCD}^4$.\footnote{A full treatment is much
more complicated, involving a range of instanton sizes, running coupling, and
temperature dependence.  All that has been lumped into the overall scale
$\Lambda$.  I've displayed the formal dependence on the coupling for later
purposes.  It makes explicit the non-perturbative character of this physics.}    So the
energy density controlled by the axion field is 
$e^{{-8\pi^2}/{g^2}} \Lambda_{\rm QCD}^4$.  The potential is minimized at
$\theta_{\rm eff.} =0$, which solves the problem we started with.    The mass$^2$
of the axion is
$e^{{-8\pi^2}/{g^2}} \Lambda_{\rm QCD}^4/F^2$.  Its interactions with matter
also scale with $\Lambda_{\rm QCD}/F$.  The failure of search experiments, so far,
together with astrophysical limits, constrain $F\gsim 10^9$ Gev.

\subsubsection{Cosmology}

Now let us consider the cosmological implications \cite{axioncosmology}.   Peccei-Quinn symmetry is
unbroken at temperatures $T\gg F$.   When this symmetry breaks the initial value of
the  phase, that is
$e^{ia/F}$, is random beyond the then-current horizon scale.  One can analyze the
fate of these fluctuations by solving the equations for a scalar field in an expanding
Universe.   

The main general results are as follows.  There is an effective cosmic viscosity, which
keeps the field frozen so long as the Hubble parameter $H \equiv \overdot R/R \gg  m$, where $R$ is the expansion factor.  
 In the opposite limit
$H \ll m$ the field undergoes lightly damped oscillations,  which result in an energy
density that decays as $\rho \propto 1/R^3$.  Which is to say, a comoving volume
contains a fixed mass.   The field can be regarded as a gas of nonrelativistic particles
(in a coherent state).   There is some additional damping at intermediate stages.    
Roughly speaking we may say that the axion field, or any scalar field in a classical
regime, behaves as an effective cosmological term for $H>>m$ and as cold dark
matter for $H\ll m$.  Inhomogeneous perturbations are frozen in
while their length-scale exceeds
$1/H$, the scale of the apparent horizon, then get damped.   

If we ignore the possibility of inflation, then there is a unique result for the cosmic
axion density, given the microscopic model.    The criterion $H \lsim m$ is satisfied
for 
$T\sim \sqrt {M_{\rm Planck}/ F} \Lambda_{\rm QCD}$.   At this point the
horizon-volume contains many horizon-volumes from the Peccei-Quinn scale, but it
is still very small, and contains only a negligible amount of energy, by current
cosmological standards.    Thus in comparing to current observations, it is
appropriate to average over the starting amplitude $a/F$ statistically.    The result
of this calculation is usually quoted in the form 
$\rho_{\rm axion} / \rho_{\rm critical} \approx F/(10^{12}~{\rm Gev})$, where
$\rho_{\rm critical}$ is the critical density to make a spatially flat  Universe, which
is also very nearly the actual density.   But in the derivation of this form the
measured value of the baryon-to-photon ratio density at  present has been used.  This is
adequate for comparing to reality, but is inappropriate for our coming purpose.   If
we don't fix the baryon-to-photon ratio, but instead demand spatial flatness,  as
suggested by inflation, then what happens for 
$F > 10^{12}$ Gev.  is that the baryon density is smaller than what we observe.   

If inflation occurs before the Peccei-Quinn transition, this analysis remains valid.  
But if inflation occurs after the transition, things are quite different.

\subsubsection{Undetermined Universe and the Anthropic Principle}

For if inflation occurs after the transition, then the patches where $a$ is
approximately homogeneous get magnified to enormous size.  Each one is far larger
than the presently observable Universe.   The observable Universe no longer
contains a fair statistical sample of
$a/F$, but some particular `accidental' value.  Of course there is still a larger
structure, which Martin Rees calls the Multiverse, over which it varies.  

Now if $F>10^{12}$ Gev, we could still be consistent with cosmological constraints
on the axion density, so long as the amplitude satisfies
$ (a/F )^2 \lsim F/( 10^{12}~{\rm Gev})$.   The actual value of $a/F$, which
controls a crucial regularity of the observable Universe, is contingent in a very
strong sense -- in fact, it is different ``elsewhere''.    By my criterion, then, the laws
of physics are not unique.  

Within this scenario, the anthropic principle is correct and appropriate.  Regions
with large values of $a/F$, so that axions by far dominate baryons, seem pretty
clearly to be inhospitable for the development of complex structures.   The axions
themselves are weakly interacting and essentially dissipationless, and they dilute the
baryons, so that these too stay dispersed.  In principle laboratory experiments could
discover axions with $F > 10^{12}$ Gev.  Then we would conclude that the vast bulk
of the Multiverse was inhospitable to intelligent life, and we would be forced to
appeal to the anthropic principle to understand the anomalously modest axion
density in our Universe.  

\subsection{Coupling Non-Uniqueness? -- The Cosmological Term}

I anticipate that many physicists will consider this answer to the topic question of this Section
a cheat, regarding the cosmic axion
density as part of initial conditions, not a law of physics.   As I discussed
semi-carefully above, I think this is a distinction without a difference.  But in any
case, by following out this example a bit further we can make the case even clearer,
and touch on another central problem of contemporary physics.  

First note that even in the case of the standard axion, with inflation after the PQ
transition, there was a stage during the evolution of the Multiverse, between
inflation and the time when $H\sim m$, when
$a/F$ was a frozen random variable, constant over each Universe.  As such it played
the role of the effective $\theta$ parameter, and also controlled a contribution to
the effective cosmological term.  By any reasonable criterion these are parameters
that appear in the laws of physics, and they were not uniquely determined.   

It is very interesting to consider extending this idea to the present day \cite{barr}.  Of course
the standard axion, connected to QCD, has long since materialized.  However it is
possible that there are other axions, connected to uniformly weak interactions, that
control much less energy, have much smaller masses, and are much more weakly
coupled.  If $m \ll H$ for the current value of $H$, which translates numerically into
$m \ll 10^{-41}~{\rm Gev}$ or 
$m \ll 10^{-60}~M_{\rm Planck}$, then the amplitude of the corresponding field is
frozen in, and its potential contributes to the effective cosmological term.

Ratcheting up the level of speculation one notch further, we can consider the
hypothesis that this is the {\it only\/} source of the observed non-vanishing
cosmological term.  To avoid confusion, let me call the axion-variant which appears
here the {\it cosmion}, and use the symbols $c$, $F_c$, etc. with the obvious
meaning.  Several attractive consequences follow.  

\begin{itemize}
\item{The magnitude of the residual cosmological term is again of the general form
${\frac12} (c/F_c)^2 e^{{-8\pi^2}/{g_c^2}} \Lambda_c^4$ for
$c/F_c \ll 1$, then saturating, but now with
$g_c$ and $\Lambda_c$ no longer tied to QCD.   This could fit the observed value, for example, with
$c/F_c \sim 1$, $\Lambda_c \sim M_{\rm Planck}$, and 
$\alpha_c \sim .01$.}
\item{The freezing criterion $H \gsim m$ translates into $F_c \gsim M_{\rm
Planck}$.  If this condition holds by a wide margin, then the value of the effective
cosmological term will remain stuck on a time-scale of order $27H^{-1} (H/m)^4$,
considerably longer than the current lifetime of the Universe.  If $F_c$ is
comparable to or less than $M_{\rm Planck}$, significant conversion of the effective
cosmological term controlled by $c$ into matter is  occurring presently.}
\item{In any case, such conversion will occur eventually.   Thus we might be able to
maintain the possibility that a fundamental explanation will fix the asymptotic value
of the cosmological term at zero.}
\item{With larger values of $\alpha_c$ and smaller values of $c/F_c$, we realize an anthropic scenario, as discussed above, but now
for dark energy instead of dark matter.}
\end{itemize}

\section{What Happens If You Keep Squeezing?}

The behavior of QCD at large density is of obvious intrinsic interest, as it provides
the answer to a child-like question, to wit:   What happens, if you keep squeezing
things harder and harder?  It is also interesting  for the description of neutron star
interiors.   We'll find an amazing answer: when you squeeze hard enough, hadronic
matter becomes a transparent (!) insulator -- like a diamond \cite{squeeze}.  

\subsection{From Too Simple, to Just Simple Enough}

Why might we hope that QCD simplifies in the limit of large density? Let's first try
the tentative assumption that things are as simple as possible, and see where it takes
us.  Thus,  assume we can neglect interactions.   Then, to start with, we'll have large
Fermi surfaces for all the quarks.    This means that the active degrees of freedom,
the excitations of quarks near the Fermi surface, have large energy and momentum.  
And so we are tempted to argue  as follows.   If an interaction between these quarks
is going to alter their distribution  significantly, it must involve a finite fractional
change in the energy-momentum.  But finite fractional changes, here, means large
absolute changes, and asymptotic freedom tells us that interactions with large
transfer of energy and momentum are rare.

Upon further consideration, however, this argument appears too quick.    For one
thing, it does not touch the gluons.   The Pauli exclusion principle, which blocks
excitation of low energy-momentum quarks, in no way constrains the gluons.   The
low energy-momentum gluons interact strongly,  and (since they were the main
problem all along) it is not obvious that going to high density  has really simplified
things much at all. 

A second difficulty appears when we recall that the Fermi surfaces of many
condensed matter systems, at low temperature,  are unstable to a pairing instability,
which drastically changes their physical properties.   This phenomenon underlies
both superconductivity and the superfluidity of Helium 3.   It arises whenever there
is an effective attraction between particles on opposite sides of the Fermi surface.   
As elucidated by Bardeen, Cooper, and Schrieffer (BCS), in theory even an arbitrarily
weak attraction can cause a drastic restructuring of the ground state.   The reason a
nominally small perturbation can have a big effect here is that we are doing
degenerate perturbation theory.  Low energy excitation of pairs of particles on
opposite sides of the Fermi surface, with total momentum zero,  all can be scattered
into one another.    By orchestrating a coherent mixture of such excitations all to pull
in the same direction, the system will discover an energetic advantage. 

In condensed matter physics the occurrence of superconductivity is a difficult and
subtle affair.   This is because the fundamental interaction between electrons is
simply electrical repulsion.   In classic superconductors an effective attraction arises
from subtle retardation effects involving phonons.  For  the cuprate
superconductors the  cause is still obscure.  

In QCD, by contrast, the occurrence of {\it color superconductivity\/} is a relatively
straightforward phenomenon.     This is because the fundamental interaction
between two quarks, unlike that between two electrons, is already attractive!    
Quarks form  triplet representations of color $SU(3)$.    A pair of quarks, in the
antisymmetric color state, form an antitriplet.    So if two quarks in this arrangement
are brought together,  the effective color charge is reduced by a factor of two
compared to when they are separated.     The color flux emerging from them is
reduced, and this means the energy in the color field is less, which implies an
attractive force.  So we should consider very carefully what color superconductivity
can do for us.   

\subsection{Consequences of Color Superconductivity}

The two central phenomena of ordinary superconductivity are the Meissner effect
and the energy gap.     The Meissner effect is the phenomenon that magnetic fields
cannot penetrate far into the body of a superconductor --  supercurrents arise to
cancel them out.   Of course, electric fields are also screened, by the motion of
charges.  Thus electromagnetic fields in general become short-ranged.   Effectively, it
appears as if  the photon has acquired a mass.   Indeed, that is just what emerges
from the equations.    We can therefore anticipate that in a color superconductor
color gluons will acquire a mass.   That is very good news, because it removes our
problem with the low energy-momentum gluons.

The energy gap means that it costs a finite amount of energy to excite electrons from
their superconducting ground state.   This is of course quite unlike what we had for
the free Fermi surface.    So  the original pairing instability, having run its course,  is
no longer present.    

With both the sensitivity to small perturbations (pairing instability) and the bad
actors (soft gluons) under control,  the remaining effects of interactions really are
small, and under good theoretical control.    We have a state of matter that is
described by weak coupling methods, but with a highly non-trivial, non-perturbative
ground state.

\subsection{Color-Flavor Locking}

The simplest and most elegant form of color superconductivity is predicted for a
slightly idealized version of real-world QCD,  in which we imagine there are exactly
three flavors of massless quarks.   (At extremely high density it is an excellent
approximation to neglect quark masses, anyway.)   Here we discover the remarkable
phenomenon of color-flavor locking.    Whereas ordinarily the symmetry among
different colors of quarks is quite distinct and separate from the symmetry among
different flavors  of quarks,  in the color-flavor locked state they become
correlated.    Both color symmetry and flavor symmetry, as separate entities are
spontaneously broken, and only a certain mixture of them survives unscathed.  

Color-flavor locking in high-density QCD drastically affects the properties of quarks
and gluons.    As we have already seen, the gluons become massive.    Due to the 
commingling of color and flavor, the electric charges of particles, which originally
depended only on their flavor, are modified.    Specifically, some of the gluons
become electrically charged, and the quark charges are shifted.   The charges of
these particles all turn out to be integer multiples of the electron's charge!      Thus
the most striking features of confinement -- absence of long-range color forces, and
integer charge for all physical excitations -- emerge as simple, rigorous consequences
of color superconductivity.    Also, since both left- and right-handed flavor
symmetries are locked to color, they are  effectively locked to one another.  Thus
chiral symmetry, which was the freedom to make independent  transformations
among the lefties and among the righties,  has been spontaneously broken.  

Altogether, there is a striking resemblance between the {\it calculated\/} properties
of the low-energy excitations in the high density limit of QCD and the {\it
expected\/} properties  --  based on phenomenological experience and models -- of
hadronic matter at moderate density.\footnote{That is, the expected behavior of
hadronic matter in the idealized world with three massless quark flavors.  The real
world of low-energy nuclear physics, in which the strange quark mass cannot be
neglected, is quite a different matter.}     This suggests the precise conjecture, that
there is no phase transition which separates them.     Unfortunately, at present both
numerical and direct experimental tests of this conjecture seem out of reach.    So it
is not quite certain that the mechanisms of confinement and chiral symmetry
breaking we find in the calculable,  high-density limit are the same as those that
operate at moderate or low density.    Still, it is astonishing  that  these properties,
which have long been regarded as mysterious and intractable,  can be demonstrated,
rigorously yet fairly simply, to occur in a physically interesting limit of QCD.

\subsection{Last Look}

Finally, it's fun to consider what the stuff looks like.   The diquark condensate is
colored, and also electrically charged, so both the original color gauge symmetry and
the original electromagnetic gauge symmetry are spontaneously broken.   However,
just as in the electroweak Standard Model both the original $SU(2)$ and the original
$U(1)$ are broken, yet a cunning combination remains to become physical
electromagnetism, so in the color-flavor locked state both color
$SU(3)$ and the original electromagnetic $U(1)$ are broken, but a cunning
combination remains valid.    This combination supports a modified `photon', part
gluon and part (original) photon, that is a massless field.   Because of the Meissner
effect and the energy gap, there are no low-energy charged excitations of these
kinds.  Some of the pseudoscalar Nambu-Goldstone bosons are charged, but they are
lifted from zero energy by the finite quark masses, that break chiral symmetry
intrinsically.    So we have an insulator.  Now because the `photon' differs a little bit
from the photon, if we shine light on a chunk of our ultradense matter, some of it
will be reflected, but most will propagate through.  So it looks like a diamond.  


\subsubsection*{Acknowledgments} I would like to thank Marty Stock for help with the manuscript.
This work is supported in part by
funds provided by the U.S. Department of Energy (D.O.E.) under cooperative research
agreement
\#DF-FC02-94ER40818.

\end{document}